\documentclass{aa}  
\usepackage{graphicx}
\usepackage{natbib}
\usepackage{txfonts}
\usepackage{array}
\usepackage{rotating}
\def\hii{H\,{\sc{ii}}}

\begin{document}
   \title{Chandra monitoring of the very massive binary WR20a \\ and the young massive cluster Westerlund\,2}

   \author{Ya\"el Naz\'e
          \inst{1}\fnmsep\thanks{Postdoctoral Researcher FNRS}
          \and
          Gregor Rauw\inst{1}\fnmsep\thanks{Research Associate FNRS}
          \and
          Jean Manfroid\inst{1}\fnmsep\thanks{Research Director FNRS}
          }

   \offprints{Y. Naz\'e}

   \institute{Institut d'Astrophysique et de G\'eophysique, Universit\'e de Li\`ege, All\'ee du 6 Ao\^ut 17 Bat. B5C, B4000-Li\`ege, Belgium\\
              \email{naze@astro.ulg.ac.be}
             }


  \abstract
   {Westerlund\,2 is a young, massive, obscured stellar cluster of our Galaxy. It harbors the most massive star with well determined parameters, WR20a (82+83M$_{\odot}$), a dozen very early O-type stars (O3--7), and a wealth of PMS stars. Although of importance, this cluster is still not well known.}  
   {The high-energy properties of this cluster, especially those of its early-type stars are examined in details. The variability of the X-ray sources is investigated. }
   {A monitoring of the field was performed using three Chandra observations. This dataset probes daily as well as monthly to yearly timescales and provides the deepest X-ray view of the cluster to date.}
   {The two Wolf-Rayet stars WR20a (WN6ha+WN6ha) and WR20b (WN6ha) were analyzed in detail. They are both very luminous and display very hard spectra, but WR20b does not seem to vary. On the contrary, WR20a, a known eclipsing, colliding-wind binary, brightens in the X-ray domain during the eclipses, i.e. when the collision is seen face-on. This can be explained by the properties of the wind-wind collision zone, whose high density leads to a large absorbing column ($2 \times 10^{24}$\,cm$^{-2}$).\\
All twelve O-type stars previously classified spectroscopically, two eclipsing binaries previously identified and nine newly identified O-type star candidates are detected in the high energy domain; ten of them could be analyzed spectroscopically. Four are overluminous, but the others present typical $L_X/L_{BOL}$ ratios, suggesting that several O-type objects are actually binaries. Variability at the $\sim2\sigma$ level was detected for a majority of the sources, of unknown origin for the putatively single objects. \\
Faint, soft, diffuse emission pervades the entire field-of-view but no clear structure can be identified, even at the position of a blister proposed to be at the origin of the TeV source HESS J1023$-$575.\\
Finally, the X-ray properties of PMS objects were also investigated, in particular the brightest flaring ones. They provided an additional argument in favor of a large distance ($\sim$8kpc) for the cluster.}
   {}

   \keywords{Galaxy: open clusters and associations: individual: Westerlund\,2 -- X-rays: stars --  stars: individual: WR20a, WR20b }

   \maketitle
%

\section{Introduction}

\subsection{The Westerlund\,2 cluster}

The young (1$-$2Myr) stellar cluster Westerlund\,2 lies in the \hii\ region RCW49. It is an intense star forming region with numerous Pre Main Sequence (PMS) objects, as enlighted by recent Spitzer data \citep[and references therein]{chu05}. The cluster also contains many early-type stars: at least two Wolf-Rayet stars, WR20a and WR20b \citep{vdH}, and twelve very hot O-stars with spectral types ranging from O3 to O6.5 \citep{rau07}. In the IR and radio domains, these stars appear surrounded by two shells, one enclosing WR20a and the core of Westerlund\,2, the other being around WR20b.

There are still many uncertainties about the properties and content of this cluster. For example, its distance is still a matter of intense debate. To cite only the most recent papers: \citet{asc07} and \citet{tsu07} proposed a `low' value of 2--5kpc, based on the X-ray and IR luminosities of PMS stars; while \citet{rau07} rather argue in favor of $8.0 \pm 1.4$\,kpc, based on the spectrophotometric properties of the O-type star population and on the eclipsing characteristics of WR20a. \citet{Dame} recently discussed the distance of the Westerlund\,2 cluster on the basis of the CO emission of the giant molecular cloud MC8 (likely associated with the cluster) and the H\,{\sc i} 21\,cm absorptions along the line of sight towards Westerlund\,2. This author found that the cluster must be located on the far side of the Carina spiral arm and inferred a kinematic distance of $6.0 \pm 1.0$\,kpc in coarse agreement with the value of \citet{rau07}. An indirect argument can be added in favor of a `high' value, using the properties of the diffuse emission detected by \citet{tow04}. The latter authors found an X-ray luminosity of about 8--$9\times10^{32}$ erg s$^{-1}$ (for an adopted distance of 2.3 kpc). Considering that the diffuse emission from massive star forming regions typically amounts to 2--$5\times10^{32}$ erg s$^{-1}$ per early-type star \citep{tow03,tow04}, we would expect values of 3--$7\times10^{33}$ erg s$^{-1}$ for a cluster containing at least 14 objects earlier than O7 (including the two Wolf-Rayet systems). This is compatible with the observed emission only if the distance is rather large (4-7kpc rather than 2.3kpc). In this paper, we will thus adopt 8kpc as the distance to Westerlund\,2.

   \begin{figure*}
   \centering
   \caption{Three-color images of Westerlund\,2 in X-rays (a, left: full field; b, right: central 8' in the middle and c, below: core of the cluster). The red, green and blue colors correspond to soft (0.3--1. keV), medium (1.--2. keV) and hard (2.--10. keV) photon energies, respectively. The images were corrected by the exposure maps and adaptively smoothed prior to RGB combination.}
              \label{3col}
    \end{figure*}
\setcounter{figure}{0}
   \begin{figure}[ht]
   \centering
   \caption{Continued.}
    \end{figure}

\subsection{X-ray properties}

Studying the X-ray emission of young clusters provides crucial information, especially if the cluster is heavily obscured as is the case of Westerlund\,2. PMS objects, mainly T Tauri stars, are more easily detected in the X-ray and IR domains, and their high-energy flares can be analysed to get their decay times, which are related to the physical size of the flaring region. In the high-mass range, the interacting colliding wind binaries and young magnetic objects are more easily identified and their properties constrained in the X-ray domain. Finally, the large-scale interaction with the surrounding medium is better studied through the analysis of the diffuse X-ray emission. 

The X-ray properties of Westerlund\,2 have been studied in increasing detail with years. The first detection of the cluster was made by the Einstein observatory: \citet{gol87} reported the presence of one X-ray emitter at the position of RCW49, interpreted as unresolved sources or truly diffuse emission (from a supernova remnant or a wind-blown bubble). Using ROSAT, \citet{bel94} detected possible diffuse emission and 24 point sources (7 of them being in our ACIS-I field-of-view), among which HD90273, MSP18\footnote{The MSP numbers come from the numbering scheme devised by \citet{mof91}.} and WR20a. The high X-ray luminosity of WR20a was suggested as a possible evidence of an on-going colliding wind phenomenon. More recently, \citet{tow04} and \citet{tsu07} reported the results of a first Chandra observation. Thanks to its high spatial resolution, this facility is the first to resolve the individual sources in the core of Westerlund\,2. More than 400 X-ray sources were identified by \citet{tsu07}, mostly PMS stars and a few early-type objects. Faint diffuse emission is also present; it extends to the south-east of the cluster and presents a soft spectrum plus a very faint hard tail \citep{tow04}.

To investigate further the high-energy properties of Westerlund\,2, especially those of its massive star population, we have obtained additional X-ray observations, which combine to provide the deepest X-ray view of the cluster. It enables us to study the variability of the X-ray sources on several timescales (short-term variations within one exposure; monthly to yearly changes between the datasets).  

The paper is organized as follows. Section 2 presents the datasets and their reduction. Sections 3 and 4 focus on the O-stars and WR-stars content, respectively. The search for a counterpart to the recently discovered TeV source and the presence of diffuse emission are discussed in section 5, while section 6 concentrates on the flaring sources. Finally, we conclude in section 7. 

\section{Observations and Data Reduction}

\subsection{Journal of observations}

Westerlund\,2 was observed three times with the ACIS-I camera onboard the Chandra X-ray Observatory, first on 2003/08/23 for 38ks (PI Garmire), then on 2006/09/05 and 2006/09/28 both for 50ks (PI Rauw), see details in Table \ref{journ}. The first dataset was taken in very faint mode, whereas the last two were taken in faint mode; in all cases, the frame time was the 3.2s standard. For all three observations, the ACIS-I chips covered a similar field-of-view but the position angle was slightly different from one exposure to another.

   \begin{table}
      \caption{Journal of observations. Phases $\phi$ refer to the ephemeris of WR20a \citep{rau07}.}
         \label{journ}
     \centering
         \begin{tabular}{lccc}
            \hline\hline
Name &  Start time & End time & $\phi$ \\
            \hline
Obs 1 & 2003/08/23, 18:21:23 & 2003/08/24, 04:53:56 & 0.34--0.46 \\
Obs 2 & 2006/09/05, 15:52:01 & 2006/09/06, 06:29:22 & 0.28--0.44 \\
Obs 3 & 2006/09/28, 05:46:20 & 2006/09/28, 20:14:43 & 0.41--0.57 \\
            \hline
         \end{tabular}
   \end{table}

The datasets were downloaded from the CXC website\footnote{http://cxc.harvard.edu/ } and we have followed their recommendations for the reduction and analysis of the data. As they had been processed by the latest version of the CIAO pipeline, we did not completely reprocess them. 

\subsection{Detailed data processing}

Background flares were only detected for the first observation, and times with background count rates larger than 0.02 cts s$^{-1}$ were therefore discarded. The astrometry was improved by comparing the position of the bright X-ray point sources WR20a and MSP18 with their counterparts in the NOMAD catalog\footnote{http://cdsarc.u-strasbg.fr/viz-bin/Cat?I/297 }. Shifts in pixels of (-0.19,0.20), (-0.10,0.04), and (0.14,0.20) were applied using the CIAO task {\sc wcs\_update} to the first, second and third observation respectively. The event lists were then reprojected without randomization, first to apply this astrometric correction, then to match the X,Y coordinates from the last observation. A merged event list was created using the CIAO task {\sc dmmerge} (with [subspace -expno] to avoid an uncorrect merging of the good time intervals). Source detection was performed on this merged event list using the CIAO wavelet algorithm. It was applied separately to 1024$\times$1024 images of the whole field (pixel of 1.4"), the central 8' (pixel of 0.5") and the core of the cluster (pixel of 0.25"). The derived source positions were used to define source/background regions for use in estimating count rates or in extracting spectra.

Exposure maps were then calculated for each observation and each energy band, assuming  a monochromatic source of energy 1 keV (total band, 0.3--10. keV), 0.7 keV (soft band S, 0.3--1.keV), 1.5 keV (medium band M, 1.--2.keV) and 3 keV (hard band H, 2.--10.keV), respectively. Merged exposure maps were obtained using the CIAO task {\sc reproject\_image}. The resulting fluxed images (in units photons cm$^{-2}$ s$^{-1}$ px$^{-2}$) were used to derive background and exposure-corrected photon fluxes\footnote{For a typical source with absorbed thermal emission ($N^{\rm H}_{avg\,West2}=10^{22}$cm$^{-2}$ and k$T$=0.3 keV), a photon flux of 1 photon cm$^{-2}$ s$^{-1}$ corresponds at the time of our observations to a count rate of 350 cts s$^{-1}$, an absorbed flux of 1.7$\times10^{-9}$ erg cm$^{-2}$ s$^{-1}$ and an unabsorbed flux of 5.7$\times10^{-8}$ erg cm$^{-2}$ s$^{-1}$.}. The merged data were also used to make the three-color images shown in Fig. \ref{3col}.

Spectrum extraction was done for each source presenting more than 100 counts in the merged event list. It was performed separately on each observation. As recommended by the CXC, we extracted the spectral response files with the tasks {\sc mkacisrmf} and {\sc mkarf} (the latter including a mask file when the considered source was close to a gap and affected by the spacecraft's dithering). When there were less than 100 counts per observation, the individual spectra were merged using the FTOOLS tasks {\sc mathpha, addrmf, addarf}. Lightcurve were extracted for the brightest sources using the CIAO task {\sc dmextract}.

Throughout our analysis, we have used CIAO 3.4, CALDB 3.3.0.1 and Xspec 11.2.

\section{O-type stars}

\subsection{Detection in the optical range}
The first spectroscopic observations of the early-type population in Westerlund\,2 revealed six O6-7 V: stars \citep{mof91}. The presence of such `late' objects seemed incompatible with the presence of a very massive binary such as WR20a (see below). A new observing campaign was therefore undertaken, which led to a serious revision of the spectral types \citep{rau07}: at least twelve O-type objects are present in Westerlund\,2, with several of them displaying a very early type (O3--4). All twelve O-type stars catalogued by \citet{rau07} are detected in the X-ray domain (see Fig. \ref{center}), but their characteristics differ significantly from one to another (see below). 

From the photometry of \citet{rau07}, we could also select O-star candidates: they have colors typical of early-type stars belonging to the cluster (B--V between 1.3 and 1.6 mag, see Fig. 7 of that paper) and absolute magnitudes smaller than $-$3.9, a value typical of O9.5V stars \citep{mar05}. Some of these candidates are also definite X-ray emitters. The optical properties of this subgroup, composed of 9 objects, are presented in Table \ref{cand}. Their color excess was derived from the observed color using an intrinsic color of $-$0.27, typical of O-type stars \citep{mar06}. Then an estimation of their spectral type was made from their absolute V magnitude at 8kpc: they are late O-type stars, O7--O9.5. The relevant bolometric correction from \citet{mar06} was then applied to find an estimate of their bolometric luminosities (Table \ref{counts}) .  

   \begin{figure*}
   \centering
   \caption{The core of the Westerlund\,2 cluster seen in visible (a, left) and X-rays (b, right). X-ray contours superimposed on the visible image (c, below; the X-ray data were adaptively smoothed). }
              \label{center}
    \end{figure*}
\setcounter{figure}{1}
   \begin{figure}[ht]
   \centering
   \caption{The core of the Westerlund\,2 cluster seen in visible (a, left) and X-rays (b, right). X-ray contours superimposed on the visible image (c, below; the X-ray data were adaptively smoothed). }
              \label{center}
    \end{figure}

We have added the O7 star HD90273 to the list of O-type objects to analyze. Note that, since it is much more luminous and less absorbed than the stars of Westerlund\,2, it is probably a foreground object, however we have calculated its luminosity as if it were an O7 supergiant at 8kpc. 

Spectral types, bolometric luminosities and absorbing columns (derived from the color excess using $N_{\rm int}^{\rm H}$=5.8$\times10^{21}\times$E(B--V), \citealt{boh78}) for the known O-type stars \citep[from][]{rau07} and the candidates objects (calculated from the photometry as explained above) are shown in Table \ref{counts}.

   \begin{table}
      \caption{Properties of candidate late-O stars. The last two lines reproduces the data of two eclipsing binaries \citep[from][]{rau07}.}
         \label{cand}
     \centering
         \begin{tabular}{lccc}
            \hline\hline
Name &  RA,DEC (J2000) & V$\pm\sigma_V$ & B--V$\pm$$\sigma_{B-V}$ \\
            \hline
MSP32  &10:24:03.78,-57:44:39.8 &15.355$\pm$0.010 &   1.285$\pm$0.014   \\
MSP120 &10:23:58.44,-57:45:13.0 &16.211$\pm$0.009 &   1.539$\pm$0.016   \\
MSP165 &10:23:55.18,-57:45:26.9 &15.604$\pm$0.009 &   1.602$\pm$0.013   \\
MSP168 &10:24:01.61,-57:45:27.8 &14.829$\pm$0.031 &   1.260$\pm$0.039   \\
MSP441 &10:24:01.45,-57:45:31.3 &14.918$\pm$0.091 &   1.283$\pm$0.138   \\
Src 1  &10:24:21.28,-57:47:27.6 &14.671$\pm$0.022 &   1.412$\pm$0.025   \\
Src 2  &10:24:16.25,-57:43:43.8 &14.914$\pm$0.015 &   1.597$\pm$0.017   \\
Src 3  &10:24:06.64,-57:47:15.8 &16.473$\pm$0.011 &   1.644$\pm$0.019   \\
Src 4  &10:24:02.28,-57:45:35.3 &13.302$\pm$0.039 &   1.371$\pm$0.059   \\
MSP44  &10:24:00.5,-57:44:45    &15.67            &   1.25              \\
MSP223 &10:23:59.2,-57:45:40    &15.75            &   1.50              \\
            \hline
         \end{tabular}
   \end{table}

   \begin{sidewaystable*}
\begin{minipage}[t]{\textwidth}
      \caption{Properties of the early-type stars in the optical and X-ray domains. Background- and exposure-corrected source fluxes (in 10$^{-5}$photons cm$^{-2}$ s$^{-1}$) are estimated for each observation from the fluxed images in a region of radius $r$ (in px) and for the 0.3--10.keV band. The last columns give the average hardness ratios and dereddened X-ray luminosity in 10$^{32}$ erg\,s$^{-1}$ and for the 0.5--10.keV energy band (see text for details). a colon (:) indicates an uncertain value.}
         \label{counts}
     \centering
\renewcommand{\footnoterule}{}
         \begin{tabular}{lcccccccccccc}
            \hline\hline
\hspace*{-0.3cm}&\\
Star &  Sp. Type & $\log(L_{BOL}/L_{\odot})$ & $N_{\rm int}^{\rm H}/10^{22}$ & $r$ & Mean ph. flux& Obs 1 & Obs 2 & Obs 3 & HR$_1$ & HR$_2$ & $L_X^{unabs}$ & $\log(L_X/L_{BOL})$\\
\hspace*{-0.3cm}&\\
            \hline
WR20a   &WN6ha+WN6ha& 6.28& 1.13& 4 & 11.87$\pm$0.17 & 10.3$\pm$0.4 & 11.2$\pm$0.3 & 13.2$\pm$0.3 & 0.64$\pm$0.03 & 0.385$\pm$0.012 & 51.7\footnote{from Table \ref{spec}} & $-$6.15\\
WR20b   & WN6ha     & 5.86& 1.04& 9 & 3.77$\pm$0.09 &  3.69$\pm$0.18 & 3.72$\pm$0.15 &  3.88$\pm$0.16  & 0.61$\pm$0.08 & 0.656$\pm$0.016 & 19.0$^a$ & $-$6.16 \\
HD90273 & O7I:      &5.80:&0.22:& 20& 0.81$\pm$0.06 &  1.03$\pm$0.11 & 0.77$\pm$0.09 &  0.66$\pm$0.10  & $-0.83\pm$0.02 & $-0.79\pm$0.24 & 2.82$^a$ & $-$6.93 \\
MSP18   & O5.5V-III & 5.89& 0.88& 3 & 6.92$\pm$0.14 &  7.4$\pm$0.4 & 6.4$\pm$0.2 &  7.4$\pm$0.2  & 0.38$\pm$0.05 & 0.19$\pm$0.02 & 35.8$^a$ & $-$5.92 \\
MSP32\footnote{OB candidate} & O9.5V:&4.61:&0.90:& 3 & 0.025$\pm$0.014 &  0.07$\pm$0.04 & 0.006$\pm$0.022 &  0.03$\pm$0.02  & $-0.15\pm$0.50 & $-0.92\pm$1.38 & 0.15& $-$7.02 \\
MSP44\footnote{eclipsing binary} & O9.5V:&4.44:&0.88:& 4 & 0.23$\pm$0.03 &  0.27$\pm$0.08 & 0.23$\pm$0.04 &  0.23$\pm$0.05  & 1. & 0.11$\pm$0.12 & 0.53& $-$6.31 \\
MSP120$^b$ & O9.5V: &4.58:&1.00:& 4 & 0.050$\pm$0.018 &  0.06$\pm$0.05 & 0.08$\pm$0.03 &  0.01$\pm$0.02  & $-0.07\pm$0.47 & $-0.47\pm$0.59 & 0.41& $-$6.54 \\
MSP151  & O7III     & 5.37& 0.95& 2 & 0.14$\pm$0.02 &  0.12$\pm$0.04 & 0.17$\pm$0.03 &  0.13$\pm$0.03  & 0.16$\pm$0.29 & $-0.28\pm$0.16 & 0.57 & $-$7.19 \\
MSP157  &O5.5V      & 5.37& 0.92& 1.5& 0.25$\pm$0.02 &  0.35$\pm$0.05 & 0.21$\pm$0.04 &  0.20$\pm$0.04  & 0.27$\pm$0.20 & $-0.10\pm$0.10 & 1.16$^a$& $-$6.89 \\
MSP165$^b$ & O7V:   &5.09:&1.08:& 4 & 0.52$\pm$0.04 &  0.48$\pm$0.06 & 0.63$\pm$0.09 &  0.45$\pm$0.08  & 0.57$\pm$0.16 & $-0.49\pm$0.09 & 1.98$^a$& $-$6.37 \\
MSP167 & O6III      & 5.14& 0.98& 2 & 2.05$\pm$0.07 &  1.80$\pm$0.12 & 2.25$\pm$0.11 &  2.00$\pm$0.13  & 0.57$\pm$0.07 & $0.00\pm$0.04 & 8.85$^a$& $-$5.77 \\
MSP168$^b$ & O8V:   &4.90:&0.89:& 2 & 0.079$\pm$0.017 &  0.05$\pm$0.03 & 0.08$\pm$0.02 &  0.11$\pm$0.02  & $-0.23\pm$0.27 & $-0.42\pm$0.25 & 0.41& $-$6.87 \\
MSP171 & O5V        & 5.58& 1.00& 4 & 0.16$\pm$0.03 &  0.13$\pm$0.03 & 0.18$\pm$0.04 &  0.13$\pm$0.09  & 0.22$\pm$0.27 & $-0.93\pm$0.35 & 1.09& $-$7.13 \\
MSP175 & O5.5V-III  & 5.49& 0.91& 2 & 0.41$\pm$0.03 &  0.34$\pm$0.05 & 0.41$\pm$0.05 &  0.45$\pm$0.05  & 0.34$\pm$0.14 & $-0.56\pm$0.08 & 1.78$^a$ & $-$6.82 \\
MSP182 & O4V-III    & 5.45& 0.93& 4 & 0.18$\pm$0.03 &  0.20$\pm$0.04 & 0.12$\pm$0.04 &  0.22$\pm$0.06  & $-0.24\pm$0.16 & $-0.76\pm$0.20 & 0.58 & $-$7.27 \\
MSP183 & O4V        & 6.07& 1.03& 2 & 0.70$\pm$0.04 &  0.67$\pm$0.07 & 0.60$\pm$0.06 &  0.86$\pm$0.09  & 0.09$\pm0.10$ & $-0.44\pm$0.06 & 4.49$^a$& $-$7.00 \\
MSP188 &O4V-III     & 5.93& 0.96& 2.5& 3.43$\pm$0.08 &  3.65$\pm$0.17 & 3.41$\pm$0.14 &  3.28$\pm$0.14  & 0.23$\pm$0.05 & $-0.10\pm$0.03 & 14.4$^a$& $-$6.35 \\
MSP199 & O3V        & 5.61& 0.97& 2 & 0.11$\pm$0.02 &  0.10$\pm$0.03 & 0.10$\pm$0.03 &  0.12$\pm$0.03  & 0.50$\pm$0.42 & $-0.56\pm$0.18 & 0.57& $-$7.44 \\
MSP203 & O6V-III    & 5.84& 0.95& 2 & 0.57$\pm$0.04 &  0.53$\pm$0.07 & 0.49$\pm$0.05 &  0.69$\pm$0.06  & 0.12$\pm$0.11 & $-0.35\pm$0.07 & 3.17$^a$ & $-$6.92 \\
MSP223$^b$ &O8.5V:  &4.80:&1.00:& 2 & 0.074$\pm$0.015 &  0.04$\pm$0.03 & 0.08$\pm$0.02 &  0.08$\pm$0.02  & 0.19$\pm$0.41 & 0.18$\pm$0.21 & 0.25& $-$6.98 \\
MSP263 & O6.5V      & 5.45& 1.09& 4 & 0.20$\pm$0.02 &  0.19$\pm$0.04 & 0.24$\pm$0.04 &  0.18$\pm$0.03  & 0.30$\pm$0.18 & $-0.47\pm$0.13 & 1.17$^a$& $-$6.96 \\
MSP441$^b$ & O8V:   &4.89:&0.90:&1.7 & 0.084$\pm$0.015 &  0.09$\pm$0.03 & 0.07$\pm$0.02 &  0.10$\pm$0.03  & 0.25$\pm$0.40 & $-0.07\pm$0.19 & 0.30& $-$7.00 \\
Src 1$^b$ & O6V:    &5.31:&0.98:& 5 & 0.071$\pm$0.015 &  0.07$\pm$0.03 & 0.10$\pm$0.03 &  0.05$\pm$0.02  & 0.15$\pm$0.32 & $-0.91\pm$0.34 & 0.43& $-$7.26 \\
Src 2$^b$ & O8.5III:&5.22:&1.08:& 6 & 0.46$\pm$0.03 &  0.54$\pm$0.07 & 0.46$\pm$0.05 &  0.40$\pm$0.05  & 0.63$\pm$0.13 & $-0.45\pm$0.08 & 1.95$^a$& $-$6.52 \\
Src 3$^b$ & O9.5V:  &4.61:&1.10:& 4 & 0.036$\pm$0.010 &  0.05$\pm$0.02 & 0.038$\pm$0.017 &  0.023$\pm$0.015  & $>0.95$ & $-0.40\pm$0.32 & 0.13& $-$7.08 \\
Src 4$^b$ & O9.5I:  &5.46:&0.95:& 1.5 & 0.146$\pm$0.018 &  0.06$\pm$0.03 & 0.25$\pm$0.04 &  0.10$\pm$0.03  & 1.02$\pm$0.21 & 0.26$\pm$0.12 & 0.57& $-$7.28 \\
            \hline
         \end{tabular}
\end{minipage}
   \end{sidewaystable*}

   \begin{table*}
      \caption{Best-fitting models and X-ray fluxes at Earth for the brightest early-type stars in Westerlund\,2. For each parameter, the lower and upper limits of the 90\% confidence interval (derived from the {\sc error} command under XSPEC) are noted as indices and exponents, respectively. The normalisation factors are defined as $10^{-14}\int n_e n_{\rm H} dV/4\pi D^2$, where $D$, $n_e$ and $n_{\rm H}$ are respectively the distance to the source (8kpc), the electron and proton density of the emitting plasma. Abundances were fixed to solar except for the WR stars (for more details, see text).}
         \label{spec}
     \centering
         \begin{tabular}{l cccccccc}
            \hline\hline
Star &  Obs & $N_{\rm int}^{\rm H}$ & $N^{\rm H}$ & k$T$ & norm & $\chi^2_{\nu}$ (dof) & $f_{\rm X}^{\rm abs}$ & $L_{\rm X}^{\rm unabs}$\\
&     & $10^{22}$~cm$^{-2}$ & $10^{22}$~cm$^{-2}$  & keV   & $10^{-3}$cm$^{-5}$         & & $10^{-13}$~erg\,cm$^{-2}$\,s$^{-1}$ & $10^{33}$~erg\,s$^{-1}$  \\
            \hline
\vspace*{-0.3cm}&\\
WR20a & Obs 1& 1.13& 1.53$_{1.20}^{1.89}$ & 0.87$_{0.24}^{1.04}$ & 1.57$_{0.98}^{16.7}$ &  & & \\
\vspace*{-0.3cm}&\\
 & & & 5.74$_{1.89}^{14.6}$ & 1.96$_{0.87}^{6.38}$ & 1.07$_{0.29}^{10.7}$ & 1.15 (33)& 4.51& 4.95\\
\vspace*{-0.3cm}&\\
WR20a & Obs 2& 1.13& 1.63$_{1.40}^{1.92}$ & 0.42$_{0.32}^{0.50}$ & 4.96$_{3.04}^{10.9}$ &  & & \\
\vspace*{-0.3cm}&\\
 & & & 5.48$_{3.92}^{7.27}$ & 1.30$_{1.16}^{1.52}$ & 2.61$_{1.69}^{3.87}$ & 1.26 (97)& 4.10& 4.88\\
\vspace*{-0.3cm}&\\
WR20a & Obs 3& 1.13& 1.93$_{1.62}^{2.24}$ & 0.35$_{0.29}^{0.43}$ & 9.60$_{4.43}^{21.6}$ & & & \\
\vspace*{-0.3cm}&\\
 & & & 3.28$_{2.40}^{4.53}$ & 1.55$_{1.39}^{1.75}$ & 1.88$_{1.44}^{2.53}$ & 1.02 (119)& 5.04& 5.69\\
\vspace*{-0.3cm}&\\
WR20b & Obs 1& 1.04 & 1.23$_{0.67}^{1.88}$ & 3.62$_{2.44}^{8.05}$ & 0.26$_{0.18}^{0.38}$ & 1.24 (24)& 2.06& 1.88\\
\vspace*{-0.3cm}&\\
WR20b & Obs 2& 1.04& 2.18$_{1.67}^{2.86}$ & 2.47$_{1.94}^{3.13}$ & 0.41$_{0.32}^{0.58}$ & 1.30 (35)& 1.97& 1.77\\
\vspace*{-0.3cm}&\\
WR20b & Obs 3& 1.04& 1.03$_{0.67}^{1.48}$ & 5.21$_{3.70}^{6.72}$ & 0.22$_{0.18}^{0.28}$ & 1.97 (38)& 2.29& 2.05\\
\vspace*{-0.3cm}&\\
MSP18 & Obs 1& 0.88& 0.$_{0.}^{0.18}$ & 3.24$_{2.53}^{4.18}$ & 0.31$_{0.28}^{0.36}$ & 0.92 (18)& 2.88& 3.70\\
\vspace*{-0.3cm}&\\
MSP18 & Obs 2& 0.88& 0.$_{0.}^{0.13}$ & 3.30$_{2.81}^{3.64}$ & 0.27$_{0.26}^{0.30}$ & 1.10 (59)& 2.54& 3.24\\
\vspace*{-0.3cm}&\\
MSP18 & Obs 3& 0.88& 0.$_{0.}^{0.11}$ & 3.04$_{2.65}^{3.43}$ & 0.32$_{0.31}^{0.36}$ & 1.00 (56)& 2.88& 3.78\\
\vspace*{-0.3cm}&\\
MSP167 & Obs 1& 0.98& 0.35$_{0.}^{0.88}$ & 1.58$_{1.07}^{2.27}$ & 0.11$_{0.08}^{0.19}$ & 1.02 (12)& 0.52& 0.77\\
\vspace*{-0.3cm}&\\
MSP167 & Obs 2& 0.98& 0.$_{0.}^{0.22}$ & 2.32$_{1.91}^{2.72}$ & 0.10$_{0.09}^{0.12}$ & 1.95 (21)& 0.72& 1.10\\
\vspace*{-0.3cm}&\\
MSP167 & Obs 3& 0.98& 0.68$_{0.}^{1.15}$ & 1.28$_{0.91}^{2.32}$ & 0.16$_{0.09}^{0.28}$ & 0.95 (11)& 0.56& 0.78\\
\vspace*{-0.3cm}&\\
MSP188 & Obs 1& 0.96& 0.45$_{0.23}^{0.69}$ & 1.22$_{1.00}^{1.40}$ & 0.24$_{0.20}^{0.33}$ & 1.56 (26)& 0.88& 1.51\\
\vspace*{-0.3cm}&\\
MSP188 & Obs 2& 0.96& 0.36$_{0.}^{0.52}$ & 1.33$_{1.20}^{1.86}$ & 0.22$_{0.16}^{0.26}$ & 0.73 (33)& 0.88& 1.49\\
\vspace*{-0.3cm}&\\
MSP188 & Obs 3& 0.96& 0.70$_{0.56}^{0.84}$ & 0.97$_{0.84}^{1.09}$ & 0.30$_{0.25}^{0.38}$ & 1.52 (30)& 0.74& 1.33\\
\vspace*{-0.3cm}&\\
HD90273 & Mgd& 0.22& 0.01$_{0.}^{0.38}$ & 0.41$_{0.22}^{0.52}$ & 0.02$_{0.}^{0.17}$ & 0.88 (33)& 0.16& 0.28\\
\vspace*{-0.3cm}&\\
MSP157 & Mgd& 0.92& 0.$_{0.}^{0.51}$ & 1.53$_{1.08}^{1.92}$ & 0.010$_{0.}^{0.018}$ & 0.97 (8)& 0.057& 0.12\\
\vspace*{-0.3cm}&\\
MSP165 & Mgd& 1.08& 1.77$_{1.29}^{2.48}$ & 0.28$_{0.20}^{0.38}$ & 1.66$_{0.}^{13.2}$ & 0.59 (10)& 0.088& 0.20\\
\vspace*{-0.3cm}&\\
MSP175 & Mgd& 0.91& 1.31$_{1.18}^{1.88}$ & 0.21$_{0.14}^{0.28}$ & 2.25$_{0.}^{59.8}$ & 1.20 (15)& 0.057& 0.18\\
\vspace*{-0.3cm}&\\
MSP183 & Mgd& 1.03& 0.56$_{0.36}^{0.85}$ & 0.63$_{0.45}^{0.82}$ & 0.10$_{0.}^{0.25}$ & 0.94 (20)& 0.14& 0.45\\
\vspace*{-0.3cm}&\\
MSP203 & Mgd& 0.95& 0.19$_{0.}^{0.47}$ & 0.99$_{0.74}^{1.14}$ & 0.03$_{0.02}^{0.05}$ & 1.09 (21)& 0.11& 0.32\\
\vspace*{-0.3cm}&\\
MSP263 & Mgd& 1.09& 1.09$_{0.62}^{1.65}$ & 0.31$_{0.18}^{0.53}$ & 0.25$_{0.}^{5.98}$ & 0.89 (7)& 0.034& 0.12\\
\vspace*{-0.3cm}&\\
Src 2  & Mgd& 1.08& 1.42$_{0.71}^{1.71}$ & 0.39$_{0.33}^{0.68}$ & 0.40$_{0.}^{0.77}$ & 1.20 (16)& 0.088& 0.19\\
\vspace*{-0.3cm}&\\
            \hline
         \end{tabular}
   \end{table*}

\subsection{X-ray properties}
We first derived the photon rates using the fluxed images for the known O-type stars, the candidates late-O and the eclipsing binaries detected in the X-ray domain (see Table \ref{counts})\footnote{Note that the values of the photon rates given in that table are raw, i.e. uncorrected for the encircled energy fraction, whereas the X-ray luminosity, derived from spectral fits, include that correction.}. The extraction region used for each star was a circle whose radius $r$, given in the second column of the table in Chandra pixels, is a compromise chosen to include most source photons while avoiding contamination from nearby sources; background regions were surrounding annuli or nearby circles devoid of sources. Errors on the photon rates and luminosities were estimated from the relative error on the net counts. Hardness ratios were calculated from the photon fluxes in the S, M, and H bands using HR$_1$=(M-S)/(M+S) and HR$_2$=(H-M)/(H+M). The X-ray luminosities in the 0.5--10.keV band were evaluated for a distance of 8 kpc using spectroscopic fits (only rough ones for the faintest objects) and were dereddened only for the interstellar column.  

The brightest objects had enough counts to provide a detailed X-ray spectrum and they were thus also analyzed spectroscopically. The derived best-fit parameters are shown in Table \ref{spec}. The second column provides the origin of the spectrum: Obs 1, 2, 3 or $Mgd$ (in case the individual spectra did not present enough counts and the individual data from the 3 observations had to be merged prior to analysis). The fitted model has the form {\tt wabs($N_{\rm int}^{\rm H}$)*wabs($N^{\rm H}$)*mekal(k$T$)} or {\tt \small wabs($N_{\rm int}^{\rm H}$)*[wabs($N^{\rm H}_1$)*mekal(k$T_1$)+wabs($N^{\rm H}_2$)*mekal(k$T_2$)]}, with {\tt wabs($N_{\rm int}^{\rm H}$)} fixed to the interstellar column \citep{rau07,boh78}. In the table, quoted fluxes and luminosities are in the 0.5$-$10.0\,keV energy range and the unabsorbed luminosities $L_{\rm X}^{\rm unabs}$ are corrected only for the interstellar absorbing column. Note that lightcurves and spectra were extracted in the same regions as defined above. An $L_X/L_{BOL}$ figure (Fig. \ref{lxlbol}) summarizes our findings. 

\subsubsection{Bright X-ray emitters}
Since the first detection of massive stars in the X-ray domain, it is known that their X-ray luminosity scales with their bolometric luminosity following $L_X\sim10^{-7}L_{BOL}$ \citep[the so-called `canonical' relation, see e.g. ][]{ber97}. Recent results have shown that this relation is rather tight for homogeneous groups of stars (NGC\,6231 see \citealt{san06} and Carina OB1 see \citealt{ant07}). Outliers are generally peculiar objects, i.e. magnetic stars or colliding-wind binaries. In Westerlund\,2, four of the bright sources clearly lie above the `canonical' relation with  $L_X/L_{BOL}$ ratios larger than $-6.4$. The brightest one, MSP18, presents a $\log (L_X/L_{BOL})$ of $-$5.9 (to be compared with the $-$6.9 of \citealt{san06}) and a high temperature (k$T>1$keV). The short-term lightcurves of MSP18, however, do not present large, significant changes (Fig. \ref{lc}) but the luminosity dropped by 15\% in the second observation (Table\ref{counts}). Due to the limited signal-to-noise ratio, it is rather difficult to judge which parameter has changed from one observation to another (flux, temperature, absorbing column?), but the high luminosity, the hardness of the spectrum, and the variability suggest the star to be a colliding-wind binary or a magnetic object. Note that binarity is also suggested by the position of the star in the optical colour-magnitude diagram of Westerlund\,2 \citep{rau07}. MSP167 and 188 present similar properties, with $\log (L_X/L_{BOL})$ of $-$5.8 and $-$6.3, respectively, and a variability at the 2$\sigma$ level between pointings. Although no large variation are seen for the full band short-term lightcurves (Fig. \ref{lc}), MSP188 exhibits some peculiar 2$\sigma$ change (an increase followed by a decrease) in its soft X-ray flux during the last observation. Finally, MSP165 also displays a rather large $\log (L_X/L_{BOL})$ of $-$6.4 and some variability (at the 2$\sigma$ level, the source is 30\% more luminous in the second observation), but the source is overall quite soft. All of these sources should be considered as potential binaries, and deserve a spectroscopic monitoring in the visible domain which is currently under way.

The other bright objects have more typical $\log (L_X/L_{BOL})$, with values between $-$6.5 and $-$7.3. Among these, MSP157 and 203 display unusually high temperatures. Some of these objects also exhibit 2--3$\sigma$ variations: HD90273 is 36\% less luminous in the last observation, MSP183 and 203 are 35\% more luminous in the last observation, the flux of MSP157 nearly doubles in the first observation, whereas that of Src2 increases by 35\%. If these stars are really single, as their modest luminosity suggests, the cause for their variability is hard to define.  

\subsubsection{Faint X-ray sources}
The faintest X-ray emitters of this sample (13 out of 26) did not have enough counts for a detailed spectroscopic analysis (less than 100 counts in total). We must therefore rely only on their photon fluxes and rough `colour' properties. In this category, we find four O-type stars classified by \citet{rau07}: from their bolometric luminosity and extinction, we would have expected at least 300 counts for each of these but we detect less than 100 counts. This is especially surprising for MSP199, an O3-4 star that should be at least as luminous as the neighbouring MSP183 (also O3-4) and certainly more luminous than the neighbouring MSP203, an O6 object. However, Fig. \ref{center} clearly shows that it is not the case. Only a few of these stars are found significantly variable at the $2\sigma$ level: MSP182 is 40\% less luminous in the second observation, MSP32 is nearly undetected and Src4 is 3--4 times more luminous in that same exposure; MSP168 is 50\% less luminous in the first observation, while this happens for Src1 in the last one. Overall, their $L_X/L_{BOL}$ follows the `canonical' relation (Fig. \ref{lxlbol}), but some of them display large HRs (MSP223, Src 4, MSP44). 

In conclusion, most O-type stars appear to follow the $L_X-L_{BOL}$ relation derived for the somewhat later O-type stars of NGC\,6231. The larger scatter for Westerlund\,2 can largely be explained by the fainter X-ray fluxes and the poorer knowledge of their properties, notably multiplicity, compared e.g. to NGC\,6231 \citep{san06}. The situation appears definitely more complicated than the faint+soft vs. bright+hard dichotomy suggested by \citet[see also Fig. \ref{lxlbol}]{tsu07}.

   \begin{figure}
   \centering
   \caption{Full band background-corrected lightcurves in units of counts per 5ks bin for the three Chandra observations (earliest at the top, latest at the bottom). Data from MSP18 are shown in the left panel, those of MSP188 in the middle one and those of WR20b in the right one. Note that the first and last bins, when affected by uncompleteness, were discarded and that the first observation of MSP18 was affected by the spacecraft's dithering (thereby explaining the lower number of recorded counts). }
              \label{lc}
    \end{figure}

   \begin{figure*}
   \centering
   \caption{Left: X-ray luminosity vs bolometric luminosity. The solid line corresponds to the canonical relation of \citet{san06}; the stars represents the Wolf-Rayet objects, the filled circles the brightest O-type stars (see Table \ref{spec}) and the open circles the faintest ones (see Table \ref{counts}). Right: Hardness ratios, symbols as in the left panel.}
              \label{lxlbol}
    \end{figure*}

\section{Wolf-Rayet stars}

\subsection{WR20a}
In 2002, WR20a was serendipitously found to be a binary with $P$=3.7d and $e$=0. Follow-up analyses revealed the system to be the most massive binary known with accurately determined masses: it consists of two very similar components of 82-83 M$_{\odot}$ each \citep[and references therein]{rau05}. This exceptional object also harbors a wind-wind collision that triggers line profile variability in the visible domain \citep{rau05}. The two components have types WN6ha, but they are only `Wolf-Rayet stars in the disguise', i.e. they are likely core-hydrogen burning stars with high mass loss rates, only slightly evolved off the main sequence.

   \begin{figure*}
   \centering
   \caption{{\it Left:} Visible lightcurve of WR20a (top figure), X-ray lightcurves with bins of 2ks (middle figure) and 5ks (for the full energy band, bottom figure). {\it Right:} Lightcurve in the medium (M, top) and hard (H, middle) bands, together with evolution of the hardness ratio $HR_2=(H-M/(H+M)$ with phase (bottom), all shown in 5ks bins. The X-ray data from the second observation are shown in solid black circles and those from the third one in open red triangles; uncomplete bins were discarded. The uncertainty on the phase $\sigma_{\phi}$ is about 0.01, taking into account the uncertainty on the period and the time elapsed since our optical observations \citep[$\sim$700d, ][]{rau07}. }
              \label{lcwr20a}
    \end{figure*}

\subsubsection{X-ray spectrum}
In \citet{rau05}, the wind composition of WR20a was analyzed: the amount of Helium was found to be consistent with solar, but the nitrogen abundance appears six times solar and the carbon abundance about or less than 4\% solar. We include these chemical peculiarities in the spectral fitting, and freeze the abundances of other elements at solar values. Two-temperature components were necessary to obtain good fits (Table \ref{spec}). Note that the low-temperature component seems the more variable from one observation to the other. 

The X-ray spectrum of WR\,20a appears overall rather hard. This is partially due to the absorption by the rather large interstellar column density, but reflects also the intrinsic hardness of the spectrum. Indeed, our two-temperature fits of the ACIS spectra of this source indicate temperatures around 1.3-2.0\,keV for the hotter component. We emphasize that plasma at temperatures that high is usually not found in the winds of single early-type stars, thus suggesting that a significant part of the X-ray emission of WR\,20a arises in the wind interaction region. Assuming that the emission arises from a radiatively cooling shock-heated plasma, we find that pre-shock velocities of about 1000\,km\,s$^{-1}$ are required to account for this hot plasma. This is twice the value of the on-axis pre-shock velocity of the wind-wind interaction that we estimated from a simple $\beta$ velocity law \citep{rau05}. This probably indicates that such a simple approximation of the pre-shock wind velocity law is not appropriate in an interacting-wind close binary system such as WR\,20a. It might also mean that the contribution from the off-axis collision, which occurs at higher speed, is not negligible. In what follows we shall thus assume that the pre-shock velocity amounts to 1000\,km\,s$^{-1}$. 

\subsubsection{X-ray lightcurve}
The three Chandra exposures were taken around the secondary eclipse: Obs1 samples the descending branch\footnote{Note that we need to be rather cautious about the results from the first observation, since during this exposure WR20a moved in and out CCD2 due to the spacecraft's dithering. }, Obs2 corresponds to phases slightly before and at the beginning of the eclipse, while Obs3 covers entirely the eclipse (Fig. \ref{lcwr20a}). A first important conclusion from the {\it Chandra} light curve of WR\,20a is the lack of an X-ray eclipse. This result can only be explained by a rather large size of the X-ray emitting region. The absence of strong variations could be explained if the X-ray emission of WR\,20a were produced in the individual winds, i.e.\ without a strong contribution from the wind interaction. However, there are two arguments against this assumption: first the rather high plasma temperature (see above) and second the X-ray luminosity that corresponds to an $L_{\rm X}/L_{\rm bol}$ ratio well in excess of that of single early-type (actually O-type) stars. Such properties are indeed more typical of colliding wind binaries (either O+O or O+WR). In the framework of a model where the X-ray emission arises mid-way between the stars as a result of a wind interaction, the absence of an eclipse sets stringent constraints on the minimum size of the X-ray emitting region. For an orbital inclination of $74.5^{\circ}$ (see \citealt{rau07} and references therein), we find that the lack of an occultation implies a strict minimum radius of the emitting region of 12\,R$_{\odot}$. 

In fact, the lightcurve not only reveals the absence of eclipse, it also indicates a brightening of the emission near conjunction phase, i.e. when the collision region is seen face-on. More precisely, on the fluxed images, WR20a appears 20-30\% more luminous in the last observation while the photon fluxes in the first two exposures are compatible within 2--3$\sigma$ (Table \ref{counts}). During each exposure, increasing trends are detected, but individual bins are generally within one or two sigma of a constant value. The last two observations, during which WR20a was not affected by dithering, follow each other in phase, and the behaviour of WR20a can thus more easily be quantified when the lightcurves are folded in phase (Fig. \ref{lcwr20a}). At the beginning, WR20a presented about 175 cnts per 5ks-bin, while at the end it reached 250 cnts per 5ks-bin. If we compare the lightcurve in the hard and medium bands\footnote{Since WR20a is heavily absorbed, there are not enough counts to make a lightcurve in the soft band.} (see Fig. \ref{lcwr20a}), the increase appears larger in the former, thereby leading to a slightly smaller hardness ratio in the last observation. The binary is thus significantly X-ray brighter and somewhat softer during the photometric eclipse. 

\subsubsection{Interpretation}
A possible explanation for this puzzling result could come from the orbital modulation of the absorption in the dense wind interaction itself. Since the winds of the two components of WR\,20a most likely have the same strength, the `stagnation' point, where the hottest plasma and hence the hardest X-ray emission is expected, should be located roughly at the centre of mass of the binary. The column density to the centre of mass is determined mostly by the density of the postshock material and by the thickness of the postshock region along the line of sight. Near quadrature phases, the column density along the line of sight is largest because the line of sight crosses the material near the centre of mass where the density is largest. 

\citet{Igor} provided a semi-analytical model for a radiative wind interaction zone\footnote{The $\chi$ cooling parameter of \citet{SBP} amounts to about $0.02$ for WR20a.}. These authors propose to calculate the cooling length $l_0$ as $l_0/r \simeq 0.01\,v^5\,r\,\sin^3{\theta}/\dot{M}$
where $r$ is the distance between the centre of the star and the stagnation point in R$_{\odot}$, $v$ is the local wind velocity in $10^3$\,km\,s$^{-1}$, $\theta$ is the angle between the tangent to the interaction front and the local wind direction and $\dot{M}$ is the mass-loss rate in $10^{-6}$\,M$_{\odot}$\,yr$^{-1}$. In the part of the interaction region around the binary axis, that is of most relevance for X-ray emission, the deflection due to the orbital motion is relatively small and hence $\theta = \pi/2$. For WR\,20a, we further adopt $r = a/2 = 27$\,R$_{\odot}$, $v = 1000$\,km\,s$^{-1}$ and $\dot{M} = 8.5 \times 10^{-6}$\,M$_{\odot}$\,yr$^{-1}$ (see \citealt{rau05} and hereabove for $v$). This then leads to $l_0 \simeq 0.86$\,R$_{\odot}$. Due to the steep temperature dependence of the radiative cooling, the width of the postshock cooling layer is of order $2\,l_0/7$ \citep{Igor}, which then yields about 0.25\,R$_{\odot}$. Antokhin et al.\ further showed that the surface density of the shock region on the binary axis $\sigma_0$ should be equal to $\sigma_0 = \dot{M}/4\,\pi\,v\,a$. For WR\,20a, this number amounts to about 0.11\,g\,cm$^{-2}$, implying a mean density in the postshock cooling layer of about $6.7 \times 10^{-12}$\,g\,cm$^{-3}$. If we assume that the wind interaction region within a radius of 12\,R$_{\odot}$ around the binary axis is delimited by two parallel planar shocks, integrating the density along the direction perpendicular to the binary axis (using equation [32] of \citealt{Igor}) and assuming the same chemical composition as \citet{rau05} yields a (highly ionized) hydrogen column density $N^{\rm H}$ of order $2 \times 10^{24}$\,cm$^{-2}$. This extremely large value\footnote{Note that due to the steep dependence on the pre-shock velocity, adopting $v = 500$\,km\,s$^{-1}$ would imply a column density about 60 times larger, and a pre-shock velocity of 1500\,km\,s$^{-1}$ a value ten times lower.} over (probably) only a limited fraction of the interaction region shows that the absorption of the X-ray emission by the material of the interaction region must indeed be quite substantial around quadrature phases. Of course, this large value of the column density only applies to a single point (the stagnation point on the binary axis). Since the X-ray emission arises over an extended part of the wind interaction region, the actual (observable) absorption will be lower. In addition, in a highly radiative wind interaction, various kinds of instabilities are expected to affect the shape of the interaction region \citep{SBP}, making it deviate significantly from a simple planar region. The latter effect will reduce the peak value of the column density towards the stagnation point, but will also broaden the episode of high absorption around quadrature phases. 

\subsection{WR20b}
Little is known on WR20b. \citet{vdH} assigned it a type WN6ha, whereas \citet{sha91} prefer WN7. As types and ages are similar for both Wolf-Rayet stars, we have used the same chemical composition for the spectral fits (Table \ref{spec}) as well as the same intrinsic color and bolometric correction as for WR20a. Combined with our photometry (V=13.51$\pm$0.02, B=14.98$\pm$0.02), this leads to a bolometric luminosity of about 7.3$\times10^5L_{\odot}$, i.e. slightly fainter than the faintest component of WR20a \citep{rau07}. 

The photon rates and hardness ratio of WR20b are presented in Table \ref{counts} while its spectral properties for each observation can be found in Table \ref{spec}. Overall, the system is 37-38\% less luminous than WR20a both in the visible and in X-rays. The spectral fits reveal a large extinction, typical of Wolf-Rayet objects, but also a high temperature. The hardness ratio is also large, even larger than for WR20a  (Fig. \ref{lxlbol}). Such properties are only  expected for interacting binaries. However, the X-ray flux appears rather constant on both short and long timescales (Fig. \ref{lc} and Table \ref{counts}), suggesting the star to be single. This contradictory result is important in the context of the current debate about the level of intrinsic X-ray emission from WN stars: while some apparently single WN stars, like WR40 \citep{gos}, remain undetected in X-rays, other objects, e.g. WR6 and 110 \citep{ski}, clearly emit X-rays. WR20b seems to belong to the latter category. An optical monitoring is under way to better constrain its nature.

\section{HESS J1023$-$575 and the diffuse emission}
\citet{aha07} recently reported on the discovery of a TeV high-energy $\gamma$-ray emission in the vicinity of Westerlund\,2 with the HESS Cherenkov telescope array. This emission appears extended with a Gaussian fit yielding $\sigma = (0.18 \pm 0.02)^{\circ}$, equivalent to a diameter of 28\,pc for a distance of 8\,kpc. The source centre lies 40s (or 5.5\arcmin) west of WR\,20a. Therefore, it is unlikely to be associated with WR\,20a alone, especially since the photon-photon absorptions in the wind interaction zone of the colliding wind binary are likely to be extremely severe, preventing a putative TeV emission from escaping from the inner regions of the wind interaction zone of WR\,20a. The size of the source is actually more typical of a wind-blown bubble. \citet{aha07} accordingly propose the TeV source to be associated with a blister-like structure detected in the radio range west of the cluster \citep{WU}.

To explain the TeV emission, \citet{Bednarek} proposed a hybrid model where both leptons and hadrons contribute to the $\gamma$-ray emission from very young open clusters. In this model, the leptons are responsible for X-ray and GeV $\gamma$-rays through synchrotron and inverse Compton radiation, whilst the hadrons produce the TeV $\gamma$-ray emission through $\pi^0$ pion decay. The electrons and the hadrons would be accelerated at the shocks resulting from the interaction of the stellar winds of the massive stars with the surrounding matter. Whilst the relativistic electrons lose their energy inside the lower density cavity created by the winds of the massive stars, the relativistic protons travel over larger distances and will only lose their energy when interacting with the dense ambient giant molecular cloud. Hence the TeV emission produced by these particles should be more extended than the X-ray and GeV emission. 

An alternative explanation has been proposed by \citet{TeV} who suggest that the TeV emission seen towards Westerlund\,2 could be the result of Lorentz-boosted MeV $\gamma$-rays emitted in the de-excitation of daughter nuclei produced by the photo-disintegration of PeV cosmic-ray nuclei in the radiation field of the early-type stars inside the cluster core. 

In some of the scenarios discussed hereabove, the TeV emission is expected to have an extended counterpart in the X-ray domain. However, in the {\it Chandra} observation, the diffuse emission pervades most of the field and appears rather featureless: our X-ray data provide no clear indication of a blister-like structure (Fig. \ref{3col}). The brightest emission is to be found to the south-east of the cluster, not towards the radio blister, and no bright X-ray point source is detected in the close vicinity of HESS J1023$-$575. Moreover, the X-ray spectrum of the diffuse emission appears mainly thermal although a faint, hard component might (or might not) be non-thermal \citep{tow04}. 


\section{Variable X-ray sources}

A photometric analysis was carried out on the 0.3-10.keV fluxed images of each observation using the standard procedures for aperture photometry available under iraf. The positions of the sources have been derived from the merged data in order to increase the signal-to-noise ratio. Apertures with radii between 1.5 to 5\arcsec were used. The relative errors were estimated from the photon statistics derived from the fluxes in the uncorrected images.

\begin{figure*}
\begin{center}
\caption{\label{PMSlc} Light curves of the eight sources that display a strong flare during one of the {\it Chandra} observations of Westerlund\,2. The sources are identified as in Table \ref{PMSspec} and the number of the observation where the flare occured is also indicated. For each panel the x-axis shows the time (in ksec) elapsed since the beginning of the observation. The light curves were binned with a time bin of 5\,ksec but note that the first and last bins might be affected by uncompleteness).} 
\end{center}
\end{figure*}

\begin{table*}
\begin{center}
\caption{Spectral analysis of the flaring PMS candidates, ordered by RA. Columns 2 and 3 yield the coordinates (J2000) of the sources, whilst the fourth column indicates the number of the observation during which the flare occured. Columns 5--9 list the best fit parameters of the spectral fits with a {\tt wabs} $\times$ {\tt mekal} model. For each parameter, the lower and upper limits of the 90\% confidence interval are noted as indices and exponents, respectively. Finally, column 10 yields the $1/e$ decay time of the flare; k$T_{\rm max}$ is the maximum plasma temperature during the flare (derived from the fitted k$T$) and $l$ is the loop half-length.\label{PMSspec}}
\begin{tabular}{r c c c c r c c c c c c}
\hline
Source & RA & DEC & Obs.\ & N$^{\rm H}$ & k$T$ & $f_{\rm X}^{\rm abs}$ & $f_{\rm X}^{\rm unabs}$ & $\chi_{\nu}^2$ (dof) & $\tau_{\rm obs}$ & k$T_{\rm max}$ & $l \times F(\zeta)$\\
\#  &    &     &  \# & ($10^{22}$\,cm$^{-2}$) &(keV) & \multicolumn{2}{c}{($10^{-14}$\,erg\,cm$^{-2}$\,s$^{-1}$)} & & (ksec) &  (keV)        & (R$_{\odot}$)       \\
\hline
\vspace*{-2mm}\\
A & 10:23:34.27 & $-57$:48:09.3 & 2 & $<$0.21 & $4.05_{2.02}^{9.93}$& 1.4& 1.4& 1.75 (4)& 2.4 \\
\vspace*{-2mm}\\
B & 10:23:57.20 & $-57$:42:59.5 & 1 & $2.39^{4.78}_{1.08}$ & $5.24^{79.9}_{1.96}$ & 8.8 & 16.1 & 0.39 (8) & 8.7 & 12.8 & 4.1\\
\vspace*{-2mm}\\
C & 10:24:00.00 & $-57$:44:21.5 & 1 & $1.35^{2.11}_{0.84}$ & $\geq 3.18$ & 4.7 & 6.2 & 0.47 (9) & 6.6 \\
\vspace*{-2mm}\\
D & 10:24:00.86 & $-57$:45:56.8 & 3 & $2.07^{2.84}_{1.27}$ & $5.93^{42.9}_{3.02}$ & 5.2 & 8.9 & 1.12 (11) & 6.2 & 14.9 & 3.2\\
\vspace*{-2mm}\\
E & 10:24:01.58 & $-57$:45:22.7 & 2 & $1.56^{2.19}_{1.07}$ & $3.91^{9.41}_{2.30}$ & 6.0 & 10.9 & 0.74 (16) & 12.0: &9.0&4.8\\
\vspace*{-2mm}\\
F & 10:24:02.05 & $-57$:44:33.0 & 1 & $1.37^{0.30}_{5.74}$& $>1.78$& 2.6& 3.4& 0.76 (1)\\
\vspace*{-2mm}\\
G & 10:24:03.21 & $-57$:45:34.5 & 2 & $1.70^{2.77}_{0.88}$ & $3.22^{15.0}_{1.57}$ & 3.0 & 6.1 & 1.66 (8) & 10.6 &7.2&3.7\\
\vspace*{-2mm}\\
H & 10:24:10.15 & $-57$:46:35.9 & 3 & $3.00^{5.07}_{1.16}$ & $2.17^{9.72}_{0.98}$ & 1.2 & 4.2 & 0.28 (5) & 10.7 &4.5&3.0\\
\vspace*{-2mm}\\
\hline
\end{tabular}
\end{center}
\end{table*}

To identify flaring sources, we have compared the photon rates of each source in the three observations. Many objects appear variable (about 6\% at the 99\% significance level), but we decided to focus on those that display a pronounced inter-observation variability ($\chi^2 \geq 100$) and are bright enough to allow for a detailed analysis of their light curve and spectrum. In this way, we have identified eight objects that display a strong flare during one of the observations (see Fig.\,\ref{PMSlc}). It should be emphasized that we are not aiming at establishing a complete census of variable sources, but rather at characterizing the properties of the brightest flaring candidates. Note that in some cases we find that the observation does not cover the entire flare but only the decay phase.  

\subsection{Flaring region}
The spectra and light curves were extracted for the flare observation over circular regions with a radius of at least 1.25\,arcsec. For isolated sources, the background was evaluated over a source-free annulus around the source region. However, in the crowded inner regions of the cluster, we rather evaluated the background over a circular, source-free, region as close as possible to the source. For the lightcurves, whenever possible, we have determined the $1/e$ decay times of the flares ($\tau_{\rm obs}$) by a least square fit of the exponential decay
(see Table\,\ref{PMSspec}). Due to the limited number of counts, the light curves had to be binned into 5\,ks bins, thus limiting our ability to determine decay times shorter than this binning time. However, all but one source actually have $\tau_{\rm obs} > 5$\,ksec. On the other hand, the spectra were fitted with an absorbed ({\tt wabs}) single temperature, solar abundance, {\tt mekal} model. Unfortunately, the sources are still too faint to extract time-resolved spectra over the duration of the flare, and the spectral analysis was therefore done for the whole exposure. The results of the fits are provided in Table\,\ref{PMSspec}. The best fit plasma temperatures are all quite high ($kT \geq 2.2$\,keV) as expected for pre-main sequence (PMS) stars during a flare. We note that the best fit column densities are larger than the interstellar column ($\sim 1 \times 10^{22}$\,cm$^{-2}$) expected from the average reddening towards the massive stars of Westerlund\,2. 
 
Correlating the positions of the flaring sources with the 2MASS point source catalogue \citep{2mass}, we find that sources A, B, D, and G have a near-IR counterpart within less than one arcsecond. However, the only one with good quality 2MASS photometry is Src A ($J = 15.47$, $H = 14.80$ and $K_s = 14.46$), the other three have highly uncertain measurements.

Assuming that the flaring plasma is confined in a closed coronal loop as is the case for solar flares, we can use the simple analytical approach proposed by \citet[for an application of the method see e.g.\ \citealt{Favata05}]{Serio} to express the loop half-length $l$ as a function of the maximum temperature at the top of the loop $T_{\rm max}$ and the thermodynamic decay time $\tau_{\rm th}$: 
$$l \sim 0.132\,\tau_{\rm th}\,\sqrt{kT_{\rm max}}$$
Here $l$ is expressed in R$_{\odot}$, $kT$ in keV and $\tau_{\rm th}$ in ksec. When the temperatures are determined from fits of ACIS CCD spectra, a correction has to be applied to the observed temperature to derive the maximum temperature at the top of the loop \citep{Favata05}:
$T_{\rm max} = 0.068\,T_{\rm fit}^{1.20}$ where the temperatures are in K. The observed decay time of the flare exceeds the intrinsic thermodynamical decay time due to the effect of heating that can extend into the decay phase of the flare. The relation between the two, $F(\zeta) = \tau_{\rm obs}/\tau_{\rm th}$, can be determined from the slope $\zeta$ of the flare decay in a $\log{T}$ -- $\log{\sqrt{EM}}$ diagram \citep{Favata05}. As mentioned above, the flaring sources discussed here are too faint to allow a time-resolved spectral analysis. Therefore, we can only determine $l \times F(\zeta)$. The corresponding values of $l \times F(\zeta)$ are 3.0 -- 4.8\,R$_{\odot}$ (see Table\,\ref{PMSspec}). If we assume that the flaring loops are freely decaying with no heating, $F(\zeta) \simeq 2$ \citep{Favata05} which actually represents the lower limit on $F(\zeta)$. Therefore, the typical loop sizes are $\leq 1.5$ -- $2.4$\,R$_{\odot}$. The flaring regions of the PMS stars in Westerlund\,2 are thus rather compact and typical of many PMS as well as active main-sequence stars that display loops anchored on the stellar photosphere with no interaction with an accretion disk. 

\subsection{Further evidence for a large distance}
\citet{tsu07} derived a loose constraint on the distance towards the cluster of 2 -- 5\,kpc, based on the median X-ray luminosity of PMS stars in the mass range 2.0 -- 2.7\,M$_{\odot}$. In fact, it seems rather unlikely that the X-ray data can provide stringent constraints on the distance. If we consider for instance the X-ray luminosity of the flaring sources, we find that their mean, absorption corrected, $L_{\rm X}$ amounts to $6.7 \times 10^{32}$\,erg\,s$^{-1}$ during the flares if we adopt a distance of 8.0\,kpc. While this is a rather high X-ray luminosity, we note that for the brightest flaring sources in NGC\,6231, \citet{pms6231} derived $L_{\rm X} \sim 3 \times 10^{32}$\,erg\,s$^{-1}$. In addition, the brightest flaring sources in the Orion Nebula Cluster observed within the COUP project display peak luminosities of $8 \times 10^{32}$\,erg\,s$^{-1}$ \citep{Favata05}. The luminosities of our flaring sources in Westerlund\,2 are thus not untypical, especially since they are probably biased towards higher values: first, we have selected only the brightest flaring objects; second, these luminosities were corrected for the full column densities derived from the spectral fits. As noted above, these column densities are larger than the values estimated from the interstellar reddening of the early-type stars in the cluster core and it could thus be that our absorption-corrected X-ray fluxes are somewhat overestimated. 

Our X-ray data might however still give hints for constraining the distance of Westerlund\,2. Indeed, a total of 202 X-ray sources have an optical counterpart at less than 1\arcsec\ in the $B\,V$ photometry of \cite{rau07} (see Fig. \ref{coulmag}). The colour-magnitude diagram of these sources indicates that most of the X-ray sources with an optical counterpart appear along the main sequence of the cluster or are foreground field stars. The latter objects actually provide a large fraction of the counterparts, a fact apparently underestimated by \citet{tsu07}. In addition, the colour-magnitude diagram shows no sign of the ZAMS turn-on. Indeed, none of our flaring sources\footnote{Note that source A falls outside the field-of-view covered by our photometric data.} seems associated with an optical source in our photometry. The optical counterparts of these sources must thus be fainter than $V = 21$. For very young clusters such as e.g.\ NGC\,6231 \citep{pms6231}, the optically brightest PMS stars that display X-ray emission are found to have absolute magnitudes of $M_V \simeq 0.5$ -- 1.5, in good agreement with the predictions of evolutionary models \citep{Siess}. For a distance of 8.0\,kpc ($DM = 14.52$) and interstellar extinction $A_V = 4.7$ -- 5.8 \citep{rau07}, the optically brightest PMS stars in Westerlund\,2 would be expected to occur somewhere in the range $V = 19.7$ -- 21.9. Were the cluster at 2kpc, the turn-on would be brighter by 3 mag and would be easily spotted in our data, and counterparts to our sources could be detected. Their absence therefore suggests a large distance for the cluster. 

\begin{figure}
\begin{center}
\caption{\label{coulmag} $(B-V, V)$ colour-magnitude diagram of the X-ray 
sources that have an optical counterpart within a correlation radius of less 
than 1\arcsec. 
The dashed lines correspond to the main-sequence between spectral types O3 and 
B2 for a distance modulus of 14.52 and reddened with $A_V = 4.68$ (leftmost 
curve) or $5.84$ (rightmost curve) and assuming $R_V = 3.1$. 
\label{coulmagX}} 
\end{center}
\end{figure}

\section{Conclusion}

In this paper, we have analyzed three Chandra observations aimed at monitoring the massive objects of Westerlund\,2. In this heavily obscured region, 24 O-type stars (identified or candidates) were detected and analyzed. Most of them display typical values of their $L_X/L_{BOL}$ ratio, though the scatter is rather large (about 60\%) due to the rather low signal-to-noise ratio of the X-ray spectrum, but also the much poorer knowledge of their bolometric luminosities and multiplicities. The most X-ray luminous objects (MSP18, 165, 167 and 188) are probably binaries and need a specific follow-up. About half of the O-type stars also exhibit variability at the 2$\sigma$ level or more, whose origin is unknown in these supposedly single objects.  

WR20a, the most massive binary, displays a hard and heavily absorbed X-ray spectrum. It brightens by about 40\% in the X-ray domain during the photometric eclipses. The X-ray emission is believed to come mostly from the wind-wind collision region, which would easily explain the presence of a high temperature in the best-fit models. If the colliding wind region is sufficiently large and opaque, a modulation will appear when the system rotates and it should then be brighter when seen `face-on', as is the case during eclipses. The other Wolf-Rayet star of the field, WR20b, is thought to be single and its flux is accordingly constant. However, the star also exhibits a high temperature and a high X-ray luminosity, and its hardness ratio is even larger than for WR20a. These latter characteristics are more typical of colliding-wind binaries.  

The photometry of the other X-ray sources was examined for each observation. Many objects appear variable, and we focused on the eight brightest, flaring ones. Their lightcurve decay time and spectral properties enabled us to infer rather compact flaring regions. The comparison between the optical and X-ray properties of the X-ray sources again suggest a large distance for Westerlund\,2. 

\begin{acknowledgements}
We acknowledge support from the Fonds National de la Recherche Scientifique (Belgium) and the PRODEX XMM and Integral contracts. We also wish to thank Drs M. De Becker, H. Sana and I. Stevens for their careful reading of the manuscript and their useful comments.
\end{acknowledgements}

\end{document}